\documentclass[12pt]{amsart}
\usepackage{amssymb,latexsym}
\usepackage{graphicx}

\begin{document}
\title[Cylindrically symmetric wormholes]
{Cylindrically symmetric wormholes}
\author{Peter K. F. Kuhfittig}
\address{Department of Mathematics\\
Milwaukee School of Engineering\\
Milwaukee, Wisconsin 53202-3109}
\date{\today}

\begin{abstract}
This  paper discusses traversable wormholes that differ slightly but 
significantly from those of the Morris-Thorne type under the assumption
of cylindrical symmetry.  The throat is a piecewise smooth cylindrical 
surface resulting in a shape function that is not differentiable at some
value.  It is proposed that the regular derivative be replaced by a 
one-sided derivative at this value.  The resulting wormhole geometry 
satisfies the weak energy condition. 
\end{abstract}

\maketitle 

PAC number(s): 04.20.Jb, 04.20.Gz

\section{Introduction}
Wormholes may be defined as handles or tunnels linking different 
universes or widely separated regions of our own Universe.  That such
wormholes may be traversable by humanoid travelers was first conjectured
by Morris and Thorne~\cite{MT88} in 1988 and has led to a flurry of
activity that has continued to the present.  For a summary of the more 
recent developments see ~\cite{LLdO03}.

The wormholes discussed by Morris and Thorne (MT) are assumed to be
spherically symmetric.  It is implicitly assumed that the throat is a 
smooth surface.  A generic feature of static wormholes, whether
spherically symmetric or not, is the violation of the weak energy 
condition.

In this paper we propose a wormhole that is different from an MT 
wormhole, starting with the assumption of cylindrical symmetry.  This
assumption allows considerably more freedom in choosing the 
metric coefficients.  

It will be shown that the ``shape function" $b=b(\rho,z)$, given in
Eq.~(\ref{E:line2}) below, must be a function of $\rho$ alone.  For
physical reasons $b=b(\rho)$, being related to the mass of the wormhole 
between any two $z$ values, has to be a continuous function of $\rho$.
Again for physical reasons, moving in the $z$ direction instead, 
$b(\rho)$ can be changed abruply at some $z$, effectively replacing one 
layer of wormhole material by another.  This results in a 
jump discontinuity at this $z$ value.  We will therefore replace the 
partial derivative by a one-sided derivative.  The main conclusion is
that for this type of wormhole, together with the use of a one-sided
derivative, the weak energy condition need not be violated.

\section{The solution}\label{S:solution}
Following Islam ~\cite{jI85}, p. 20, a cylindrically symmetric static
metric can be put into the form
\begin{equation}\label{E:line1}
   ds^2=-fdt^2+l\rho^2d\theta^2+Ad\rho^2+2Bd\rho dz +Cdz^2
\end{equation}
(employing the Lorentzian signature), where $f, l, A, B$, and $C$
are all functions of $\rho$ and $z$.  Using the transformation 
$\rho'=F(\rho, z)$ and $z'=G(\rho,z)$, we get the following form 
by finding $d\rho'$ and $dz'$, solving for $d\rho$ and $dz$, and
substituting in Eq.~(\ref{E:line1}):
\begin{multline*}
   ds^2=-fdt^2+l\rho^2d\theta^2 +J^{-2}\left\{(AG_2^2-2BG_1G_2+
    CG_1^2)(d\rho')^2\right.\\
     +2\left[-AG_2F_2+B(G_2F_1+G_1F_2)-CG_1F_1\right]d\rho'dz' \\
       +\left. (AF_2^2-2BF_1F_2+CF_1^2)(dz')^2\right\},
\end{multline*}
where the subscripts denote partial derivatives (e.g., 
$F_1=\partial F/\partial\rho$) and $J=F_1G_2-F_2G_1$ is the 
Jacobian of the transformation. Since the functions $F$ and $G$ 
are arbitrary, we can let
\[
    -AG_2F_2+B(G_2F_1+G_1F_2)-CG_1F_1=0
\]
and
\[
    J^{-2}(AF_2^2-2BF_1F_2+CF_1^2)=l.
\]
Assuming there exists a nontrivial solution, the metric can be 
written in the form (omitting primes)
\[
   ds^2=-fdt^2+H(\rho,z)d\rho^2+l(\rho^2d\theta^2+dz^2).
\]
In the spirit of Morris and Thorne ~\cite{MT88}, we will write
the metric in the form 
\begin{equation}\label{E:line2}
   ds^2=-e^{2\Phi(\rho,z)}dt^2+\frac{1}{1-b(\rho,z)/\rho}d\rho^2
      +\left[K(\rho,z)\right]^2(\rho^2d\theta^2+dz^2),
\end{equation}
where $\Phi=\Phi(\rho,z)$ is called the \emph{redshift function} 
and $b=b(\rho,z)$ the
\emph{shape function}. By the usual assumption of asymptotic flatness
we require that $\Phi=\Phi(\rho,z)$ as well as its partial derivatives
approach zero as $\rho,z\rightarrow \infty$.  The function $K(\rho,z)$ 
is a positive nondecreasing function of $\rho$ that determines the 
proper radial distance from the origin, so that $(\partial/\partial
\rho)K(\rho,z)>0$.  Thus $2\pi\rho K(\rho,z)$ is the proper 
circumference of the circle passing through $(\rho,z)$.  (A similar
function appears in the metric describing a rotating wormhole \cite
{eT98,pK03}.)  Apart from these requirements we assume that $\Phi$, 
$b$, and $K$ can be freely assigned to meet the desired physical 
requirements of the wormhole.

To see how to best interpret the shape function $b=b(\rho,z)$, we
need to calculate the nonzero components of the Einstein tensor in
the orthonormal frame.  These are listed next:
\begin{multline}
   G_{\hat{t}\hat{t}}=\left(\frac{1}{\rho^2K(\rho,z)}
      \frac{\partial K(\rho,z)}{\partial\rho}+\frac{1}{2\rho^3}
      \right)\left(\rho\frac{\partial b(\rho,z)}{\partial\rho}
         -b(\rho,z)\right)\\
  +\frac{1}{[K(\rho,z)]^4}\left(\frac{\partial K(\rho,z)}
      {\partial z}\right)^2-\frac{1}{[K(\rho,z)]^3}
                \frac{\partial^2K(\rho,z)}{\partial z^2}\\
    +\left(1-\frac{b(\rho,z)}{\rho}\right)\times\\
       \left[-\frac{3}{\rho K(\rho,z)}\frac{\partial K(\rho,z)}
       {\partial\rho}-\frac{2}{K(\rho,z)}\frac{\partial^2
                K(\rho,z)}{\partial\rho^2}-\frac{1}{[K(\rho,z)]^2}
                   \left(\frac{\partial K(\rho,z)}{\partial\rho}
                         \right)^2\right]\\
   -\frac{3}{4\rho^2[K(\rho,z)]^2}\left(1-\frac{b(\rho,z)}
       {\rho}\right)^{-2}\left(\frac{\partial b(\rho,z)}
         {\partial z}\right)^2\\
    -\frac{1}{2\rho[K(\rho,z)]^2}\left(1-\frac{b(\rho,z)}{\rho}
        \right)^{-1}\frac{\partial^2b(\rho,z)}{\partial z^2},
\end{multline}
\begin{multline}
   G_{\hat{\rho}\hat{\rho}}=\frac{1}{[K(\rho,z)]^2}\left[\frac
    {\partial^2\Phi(\rho,z)}{\partial z^2}+
          \left(\frac{\partial\Phi(\rho,z)}{\partial z}\right)^2
               \right]\\
     -\frac{1}{[K(\rho,z)]^4}\left(\frac{\partial K(\rho,z)}
       {\partial z}\right)^2+\frac{1}{[K(\rho,z)]^3}
                \frac{\partial^2 K(\rho,z)}{\partial z^2}\\
   +\left(1-\frac{b(\rho,z)}{\rho}\right)\times\\
    \left[\frac{2}{K(\rho,z)}\frac{\partial\Phi(\rho,z)}
      {\partial\rho}\frac{\partial K(\rho,z)}{\partial\rho}
       +\frac{1}{\rho}\frac{\partial\Phi(\rho,z)}{\partial\rho}
             \right.\\
        +\left.\frac{1}{[K(\rho,z)]^2}\left(\frac{\partial K(\rho,z)}
             {\partial\rho}\right)^2+\frac{1}{\rho K(\rho,z)}
               \frac{\partial K(\rho,z)}{\partial\rho}\right],
\end{multline}
\begin{multline}
  G_{\hat{\theta}\hat{\theta}}=\left(-\frac{1}{2\rho^2K(\rho,z)}
    \frac{\partial K(\rho,z)}{\partial\rho}-\frac{1}{2\rho^2}
    \frac{\partial\Phi(\rho,z)}{\partial\rho}\right)
     \left(\rho\frac{\partial b(\rho,z)}{\partial\rho}-b(\rho,z)
        \right)\\
      -\frac{1}{[K(\rho,z)]^3}\frac{\partial\Phi(\rho,z)}
          {\partial z}\frac{\partial K(\rho,z)}{\partial z}
    +\frac{1}{[K(\rho,z)]^2}\left[\frac{\partial^2\Phi(\rho,z)}
       {\partial z^2}+\left(\frac{\partial\Phi(\rho,z}
              {\partial z}\right)^2\right]\\
  +\left(1-\frac{b(\rho,z)}{\rho}\right)\times
     \left[\frac{\partial^2\Phi(\rho,z)}{\partial\rho^2}
      +\left(\frac{\partial\Phi(\rho,z)}{\partial\rho}\right)^2
              \right.\\ 
    \left.\phantom{\left(\frac{\partial\Phi}{\partial\rho}\right)^2}
       +\frac{1}{K(\rho,z)}\frac{\partial\Phi(\rho,z)}{\partial\rho}
         \frac{\partial K(\rho,z)}{\partial\rho} +\frac{1}{K(\rho,z)}
            \frac{\partial^2K(\rho,z)}{\partial\rho^2}\right]\\
     +\frac{3}{4\rho^2[K(\rho,z)]^2}
        \left(1-\frac{b(\rho,z)}{\rho}\right)^{-2}
          \left(\frac{\partial b(\rho,z)}{\partial z}\right)^2\\
      +\frac{1}{2\rho[K(\rho,z)]^2}\left(1-\frac{b(\rho,z)}{\rho}
          \right)^{-1}\frac{\partial\Phi(\rho,z)}{\partial z}
              \frac{\partial b(\rho,z)}{\partial z}\\
      -\frac{1}{2\rho[K(\rho,z)]^3}\left(1-\frac{b(\rho,z)}
          {\rho}\right)^{-1}\frac{\partial K(\rho,z)}{\partial z}
              \frac{\partial b(\rho,z)}{\partial z}\\
    +\frac{1}{2\rho[K(\rho,z)]^2}\left(1-\frac{b(\rho,z)}{\rho}
        \right)^{-1}\frac{\partial^2 b(\rho,z)}{\partial z^2},
\end{multline}
\begin{multline}
   G_{\hat{z}\hat{z}}=\\ \left(-\frac{1}{2\rho^2}
     \frac{\partial\Phi(\rho,z)}{\partial\rho}-\frac{1}
     {2\rho^2 K(\rho,z)}\frac{\partial K(\rho,z)}{\partial\rho}
        -\frac{1}{2\rho^3}\right)\left(\rho\frac{\partial b(\rho,z)}
            {\partial\rho}-b(\rho,z)\right)\\
    +\frac{1}{[K(\rho,z)]^3}\frac{\partial\Phi(\rho,z)}{\partial z}
            \frac{\partial K(\rho,z)}{\partial z}\\
    +\left(1-\frac{b(\rho,z)}{\rho}\right)\times
    \left[\frac{\partial^2\Phi(\rho,z)}{\partial\rho^2}
       +\left(\frac{\partial\Phi(\rho,z)}{\partial\rho}\right)^2
           \right.\\ \left.
      +\frac{1}{K(\rho,z)}\frac{\partial\Phi(\rho,z)}{\partial\rho}
       \frac{\partial K(\rho,z)}{\partial\rho}+\frac{1}{\rho}
      \frac{\partial\Phi(\rho,z)}{\partial\rho}+\frac{2}{\rho K(\rho,z)}
           \frac{\partial K(\rho,z)}{\partial\rho}\right.\\
     \left.\phantom{\left(\frac{\partial\Phi}{\partial\rho}\right)^2}
            +\frac{1}{K(\rho,z)}\frac{\partial^2 K(\rho,z)}
     {\partial\rho^2}\right]\\
      +\frac{1}{2\rho[K(\rho,z)]^3}\left(1-\frac{b(\rho,z)}
          {\rho}\right)^{-1}\frac{\partial K(\rho,z)}{\partial z}
            \frac{\partial b(\rho,z)}{\partial z}\\
    +\frac{1}{2\rho[K(\rho,z)]^2}
         \left(1-\frac{b(\rho,z)}{\rho}\right)^{-1}
      \frac{\partial\Phi(\rho,z)}{\partial z}
           \frac{\partial b(\rho,z)}{\partial z},
\end{multline}     
\begin{multline}
   G_{\hat{\rho}\hat{z}}=\left(1-\frac{b(\rho,z)}{\rho}\right)
        ^{1/2}\times\\
  \left[\frac{1}{[K(\rho,z)]^2}\frac{\partial K(\rho,z)}{\partial\rho}
   \frac{\partial\Phi(\rho,z)}{\partial z}-\frac{1}{K(\rho,z)}
      \frac{\partial^2\Phi(\rho,z)}{\partial\rho\partial z}\right.\\
      \left.-\frac{1}{K(\rho,z)}\frac{\partial\Phi(\rho,z)}{\partial z}
        \frac{\partial\Phi(\rho,z)}{\partial\rho}
       -\frac{1}{[K(\rho,z)]^2}\frac{\partial^2 K(\rho,z)}
              {\partial\rho\partial z}\right.\\
     \left.-\frac{1}{\rho [K(\rho,z)]^2}\frac{\partial K(\rho,z)}
     {\partial z}+\frac{1}{[K(\rho,z)]^3}\frac{\partial K(\rho,z)}
        {\partial\rho}\frac{\partial K(\rho,z)}{\partial z}\right]\\
   +\left(1-\frac{b(\rho,z)}{\rho}\right)^{-1/2}\times
     \left[\frac{1}{2\rho K(\rho,z)}\frac{\partial\Phi(\rho,z)}
     {\partial\rho}\frac{\partial b(\rho,z)}{\partial z}\right.\\
      +\left.\frac{1}{2\rho[K(\rho,z)]^2}
        \frac{\partial K(\rho,z)}{\partial\rho}
         \frac{\partial b(\rho,z)}{\partial z}
           +\frac{1}{2\rho^2K(\rho,z)}
             \frac{\partial b(\rho,z)}{\partial z}\right].
 \end{multline}

We would like $b=b(\rho,z)$ to correspond to the throat of the 
wormhole for some $\rho$ and $z$.  But in that case the fraction
$1/(1-b(\rho,z)/\rho)$ is undefined at the throat.  In particular, 
$T_{\hat{t}\hat{t}}=\rho=(1/8\pi)G_{\hat{t}\hat{t}}$ is undefined.  
A remarkable feature of the solution is the following: 
the expressions
\[
   \frac{1}{1-b(\rho,z)/\rho}\qquad\text{and}\qquad\frac{\partial
           b(\rho,z)}{\partial z}
\]
always occur together.  So we must require that $\partial b(\rho,z)/
\partial z=0$, i.e., $b$ must be independent of $z$.  Accordingly, 
we will write $b=b(\rho)$ from now on and omit all terms containing 
the factor $\partial b(\rho,z)/\partial z$.

These observations allow us to interpret $b=b(\rho)$ in terms of the
usual embedding diagram, such as Fig. 1 in MT~\cite{MT88}, 
by letting $t$ be a fixed moment in time in Eq.~(\ref{E:line2}) 
and then choosing the slice $z=0$.  In MT
every circle represents a sphere because of the assumption of 
spherical symmetry.  In our case, every circle represents a cylinder,
so that the throat, or part of the throat, is a cylindrical surface
with minimal radius $\rho=\rho_0$ extending along the $z$ axis.  Just
how far the throat extends depends on the energy conditions, 
discussed in the next section. 

\section{The weak energy condition}\label{S:WEC}
Recall that the weak energy condition (WEC) can be stated as
$T_{\hat{\alpha}\hat{\beta}}\mu^{\hat{\alpha}}\mu^{\hat{\beta}}\ge0$
for all timelike and, by continuity, all null vectors and where
$T_{\hat{\alpha}\hat{\beta}}$ are the components of the stress-energy 
tensor in the orthonormal frame.

Since the stress-energy tensor has the same algebraic structure as the 
Einstein tensor, we have (from the Einstein field equations $8\pi
T_{\hat{\alpha}\hat{\beta}}=G_{\hat{\alpha}\hat{\beta}}$)
\[
\begin{array}{ccc}
    T_{\hat{t}\hat{t}}=\rho=\frac{1}{8\pi}G_{\hat{t}\hat{t}},
        & T_{\hat{\rho}\hat{\rho}}=-\tau=
                  \frac{1}{8\pi}G_{\hat{\rho}\hat{\rho}},\\
    T_{\hat{\theta}\hat{\theta}}=\frac{1}{8\pi}
       G_{\hat{\theta}\hat{\theta}},&T_{\hat{z}\hat{z}}=
          \frac{1}{8\pi}G_{\hat{z}\hat{z}},
       &T_{\hat{\rho}\hat{ z}}=\frac{1}{8\pi}G_{\hat{\rho}\hat{z}}.
\end{array}
\]
For the case of a diagonal stress-energy tensor the WEC can be 
written~\cite{KS96}
\begin{equation}
   G_{\hat{t}\hat{t}}\ge0,\quad G_{\hat{t}\hat{t}}+
      G_{\hat{\rho}\hat{\rho}}\ge0,\quad G_{\hat{t}\hat{t}}+
       G_{\hat{\theta}\hat{\theta}}\ge0, \quad
    G_{\hat{t}\hat{t}}+G_{\hat{z}\hat{z}}\ge0
\end{equation}
corresponding to the timelike vector $(1,0,0,0)$ and the null vectors
$(1,1,0,0)$, $(1,0,1,0)$, and $(1,0,0,1)$, respectively.  Because of the 
off-diagonal element $T_{\hat{\rho}\hat{z}}$, we also need to consider 
the null vector 
\[
          (1, \frac{1}{\sqrt{2}},0, \frac{1}{\sqrt{2}}), 
\]
which yields
\begin{equation}\label{E:lastWEC}
   \frac{1}{2}(G_{\hat{t}\hat{t}}+G_{\hat{\rho}\hat{\rho}})+
     \frac{1}{2}(G_{\hat{t}\hat{t}}+G_{\hat{z}\hat{z}})
         +G_{\hat{\rho}\hat{z}}\ge0.
\end{equation}

Suppose we start the investigation with the second energy condition:
\begin{multline}\label{E:second}
   G_{\hat{t}\hat{t}}+G_{\hat{\rho}\hat{\rho}}=\\
   \left(\frac{1}{\rho^2K(\rho,z)}\frac{\partial K(\rho,z)}
      {\partial\rho}\right)\left(\rho\frac{db(\rho)}{d\rho}
      -b(\rho)\right)+\frac{1}{2\rho^3}\left(\rho
                \frac{db(\rho)}{d\rho}-b(\rho)\right)\\
     +\frac{1}{[K(\rho,z)]^2}\left[\frac{\partial^2\Phi(\rho,z)}
     {\partial z^2}+\left(\frac{\partial\Phi(\rho,z)}{\partial z}
        \right)^2\right]\\
   +\left(1-\frac{b(\rho)}{\rho}\right)\times\left[\frac{2}
     {K(\rho,z)}\frac{\partial\Phi(\rho,z)}{\partial\rho}
  \frac{\partial K(\rho,z)}{\partial\rho}+\frac{1}{\rho}
     \frac{\partial\Phi(\rho,z)}{\partial\rho}\right.\\
    -\frac{2}{\rho K(\rho,z)}\frac{\partial K(\rho,z)}
       {\partial\rho}
         \left.-\frac{2}{K(\rho,z)}
             \frac{\partial^2 K(\rho,z)}{\partial\rho^2}\right];
\end{multline}
(recall that $\partial b(\rho,z)/\partial z=0.$)

The factor
\begin{equation}\label{negativeenergyexpression} 
    \rho\, db(\rho)/d\rho-b(\rho)
\end{equation} 
is negative at the throat, 
since $b(\rho_0)=\rho_0$ and $b'(\rho_0)<1.$  It follows that the second term
\[
    \frac{1}{2\rho^3}\left(\rho\frac{db(\rho)}{d\rho}-b(\rho)\right)
\]
is negative with an absolute value equal to a multiple of $1/\rho_0^2$.
The first term has a smaller absolute value, as we will see below 
(Expression~(\ref{E:firstpart2})).

At the throat, $1-b(\rho)/\rho=0.$  Since the partial derivatives of 
$\Phi=\Phi(\rho,z)$ go to zero as $z\rightarrow\infty$, $G_{\hat{t}
\hat{t}}+G_{\hat{\rho}\hat{\rho}}$ eventually becomes negative.  So 
we need to cut off the wormhole region at some $z$ value above the 
equatorial plane $z=0$.  (Since the analysis for the lower region would
follow similar lines, we will concentrate exclusively on the upper 
region.)  For proper choices of $\Phi(\rho,z)$ and 
$K(\rho,z)$ it should be possible to make the sum of the first three
terms in Eq.~(\ref{E:second}) nonnegative below this $z$ value.
Even though we have great 
freedom in choosing $\Phi$ and $K$, the existence of such functions 
cannot be taken for granted.  So let us consider the following forms:
\[
  \Phi(\rho,z)=A(\rho)G(z)\quad\text{and}\quad K(\rho,z)
      = 1+\frac{F(z)}{\rho^a},
\]
where $0<a\ll 1$.  
\begin{figure}[htbp]
\begin{center}
\includegraphics[width=4in]{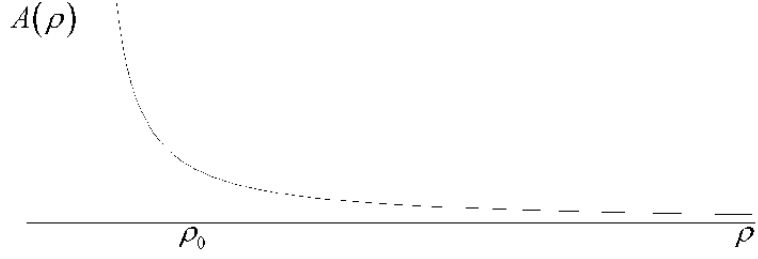}
\end{center}
\caption{\label{fig:figure1}Graph showing the qualitative features of $A(\rho)$}
\end{figure} 
\begin{figure}[htbp]
\begin{center}
\includegraphics[width=4in]{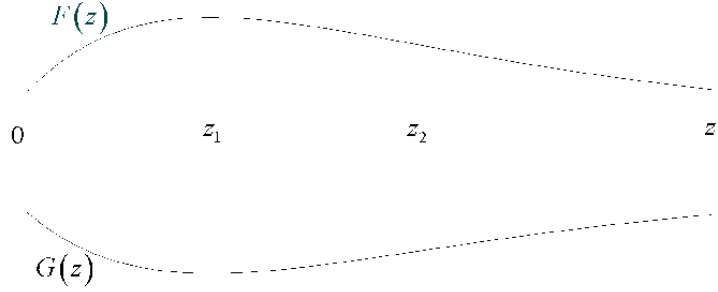}
\end{center}
\caption{\label{fig:figure2}Qualitative description of $G(z)$ and $F(z)$}
\end{figure} 
Qualitatively, the basic shapes are given in 
Figures \ref{fig:figure1} and \ref{fig:figure2}: 
$A(\rho)$ is concave up everywhere with a vertical
asymptote at $\rho=\epsilon >0$, well to the left of $\rho=\rho_0$, 
$F(z)$ is concave down and $G(z)$ concave up on an interval that 
extends well beyond $z=z_1$, where $F'(z_1)=G'(z_1)=0$, to some 
$z=z_2$.  Both $\epsilon$ and $z_2$ are to be determined later.

The partial derivatives are:
\[
    \frac{\partial\Phi(\rho,z)}{\partial\rho}=A'(\rho)G(z)>0 \qquad
     \text{for}\quad 0<z<z_2,
\]
\[
    \frac{\partial^2\Phi(\rho,z)}{\partial\rho^2}=A''(\rho)G(z)<0
     \qquad \text{for}\quad 0<z<z_2,
\]
\[
   \frac{\partial\Phi(\rho,z)}{\partial z}=A(\rho)G'(z)<0  \qquad
      \text{for}\quad 0<z<z_1,
\]
\[
   \frac{\partial^2\Phi(\rho,z)}{\partial z^2}=A(\rho)G''(z)>0\qquad
    \text{for}\quad 0<z<z_2,
\]
\[
  \frac{\partial K(\rho,z)}{\partial\rho}=-\frac{aF(z)}{\rho^{a+1}}<0
   \qquad \text{for}\quad 0<z<z_2,
\]
\[
     \frac{\partial^2 K(\rho,z)}{\partial\rho^2}=
      \frac{a(a+1)F(z)}{\rho^{a+2}}>0\qquad
     \text{for}\quad 0<z<z_2,
\]
\[
 \frac{\partial K(\rho,z)}{\partial z}=\frac{F'(z)}{\rho^{a}}>0
    \qquad \text{for} \quad 0<z<z_1,
\]
and
\[ 
    \frac{\partial^2 K(\rho,z)}{\partial z^2}=
           \frac{F''(z)}{\rho^{a}}<0\qquad
    \text{for}\quad 0<z<z_2.
\]
On the interval $(z_1,z_2)$, $G'(z)$ and $F'(z)$ change signs.

On the interval $(0, z_2)$ it seems sufficient for now to choose 
$A$ so that
\begin{multline}\label{E:chooseAG}
  \frac{1}{[K(\rho,z)]^2}\left[\frac{\partial^2\Phi(\rho,z)}
     {\partial z^2}+\left(\frac{\partial\Phi(\rho,z)}
     {\partial z}\right)^2\right]=\\
   \frac{1}{\left(1+\frac{F(z)}{\rho^a}\right)^2}
    \left\{A(\rho)G''(z)+\left[A(\rho)G'(z)\right]^2\right\}
\end{multline}
exceeds the first two terms (of order $1/\rho_0^2$)  
in Eq.~(\ref{E:second}).  This can always be accomplished by ``raising" $A(\rho)$ 
sufficiently, independently of $G$ and $F$.

As already noted, $1-b(\rho)/\rho$ is zero at the throat.  In some 
wormhole solutions~\cite{aD01,pK02} the violation actually occurs near
the throat, rather than at the throat.  For the second part of the 
right-hand side of Eq.~(\ref{E:second}), we get
\begin{multline}\label{E:secondcontinued}
  \left(1-\frac{b(\rho)}{\rho}\right)\left[-\frac{2a}{\rho}
    \frac{\frac{F(z)}{\rho^a}}{1+\frac{F(z)}{\rho^a}}
      A'(\rho)G(z)\right.\\
     \left.\phantom{\frac{F(z)}{\left(1+\frac{F(z)}{\rho^a}\right)}}
     +\frac{1}{\rho}A'(\rho)G(z)-\frac{2a^2}{\rho^2}\frac{\frac{F(z)}
       {\rho^a}}{1+\frac{F(z)}{\rho^a}}\right].
\end{multline}
As long as $a$ is small, the positive second term dominates 
on the interval $(0, z_2)$ for any
$F(z)$ and for a wide variety of choices of $A$ and $G$.  (We will 
see later that $A'(\rho)G(z)$ has to be relatively large to begin 
with.)

For the first energy condition, $G_{\hat{t}\hat{t}}\ge0$, we need to
specify $K(\rho,z)=1+F(z)/\rho^a$ more precisely: the first part of 
$G_{\hat{t}\hat{t}}$,
\begin{multline}\label{E:firstpart1}
  \left(\frac{1}{\rho^2 K(\rho,z)}\frac{\partial K(\rho,z)}
    {\partial\rho}+\frac{1}{2\rho^3}\right)\left(\rho\frac{db(\rho)}
    {d\rho}-b(\rho)\right)\\
  +\frac{1}{\left(1+\frac{F(z)}{\rho^a}\right)^4}
    \left(\frac{F'(z)}{\rho^a}\right)^2-\frac{1}
      {\left(1+\frac{F(z)}{\rho^a}\right)^3}
         \frac{F''(z)}{\rho^a}
\end{multline}
must be nonnegative on the interval $(0,z_2)$.  (Observe that the 
last two terms are positive on this interval.)  That such a function 
can be constructed can be seen by starting with the rough straight-line 
approximation for $F(z)$, the left part extending from $(0,b)$ to a 
peak at $(z_1,b+c)$: 
\[
    K(\rho,z)=1+\frac{(c/ z_1)z+b}{\rho^a}.
\]
Expression (\ref{E:firstpart1}) becomes
\begin{multline}\label{E:firstpart2}
   \left(-\frac{a}{\rho^3}\frac{\frac{F(z)}{\rho^a}}{1+\frac{F(z)}{\rho^a}}
    +\frac{1}{2\rho^3}\right)\left(\rho\frac{db(\rho)}{d\rho}-b(\rho)\right)
   +\frac{1}{\left(1+\frac{(c/ z_1)z+b}{\rho^a}\right)^4}
      \left(\frac{c/ z_1}{\rho^a}\right)^2
\end{multline}  
The first term is of order $1/ \rho_0^2$, as we saw earlier.  
The second term is smallest at $z=z_1$.  To match the
first term, we need (at least roughly)
\[
  \frac{1}{\left(1+\frac{c+b}{\rho^a}\right)^2}\frac{c/ z_1}{\rho^a}=
     \frac{1}{\rho_0}.
\]
This is a quadratic equation in $c$ with infinitely many real solutions
in terms of the other parameters.  Finally, the corner at $z=z_1$ can be
replaced by a small arc with an arbitrarily large curvature $\kappa$, where
$\kappa=F''(z)/ \left[1+(F'(z))^2 \right]^{3/2}\approx F''(z)$ near $z=z_1$.

Away from the throat we obtain the remaining part of $G_{\hat{t}\hat{t}}$
by returning to $K(\rho,t)=1+F(z)/\rho^a$ and recalling that 
$\partial b(\rho,z)/\partial z=0$:
\[
  \left(1-\frac{b(\rho)}{\rho}\right)\frac{\frac{F(z)}{\rho^a}}
    {1+\frac{F(z)}{\rho^a}}\left(\frac{a-2a^2}{\rho^2}-\frac
    {a^2\frac{F(z)}{\rho^a}}{\left(1+\frac{F(z)}{\rho^a}\right)\rho^{2}}\right).
\] 
For small $a$, this expressions is positive for \emph{any} $F(z)$.

\section{The remaining energy conditions}
The energy conditions $G_{\hat{t}\hat{t}}+
G_{\hat{\theta}\hat{\theta}}\ge0$ and $G_{\hat{t}\hat{t}}+
G_{\hat{z}\hat{z}}\ge0$ can be checked in a similar way with the understanding 
that some refinements may still have to be made.  For example, in the condition
$G_{\hat{t}\hat{t}}+G_{\hat{z}\hat{z}}\ge0$ the terms involving the ``negative
energy expression"(\ref{negativeenergyexpression}), $\rho\,db(\rho)/
d\rho-b(\rho)$, are actually positive.  Unfortunately, there is also the
dangerous negative term
\[
   \frac{1}{[K(\rho,z)]^3}\frac{\partial\Phi(\rho,z)}{\partial z}
      \frac{\partial K(\rho,z)}{\partial z}
\]
on the entire interval $(0,z_2)$ since $G'(z)$ and $F'(z)$ have opposite
signs. (In the condition $G_{\hat{t}\hat{t}}+G_{\hat{\theta}\hat{\theta}}\ge0$ 
this term has the opposite sign, so that this problem does not arise.)  
The terms not involving $1-b(\rho)/\rho$ are
\begin{multline}\label{E:fourth}
     \left(\frac{1}{2\rho^2K(\rho,z)}\frac{\partial K(\rho,z)}{\partial\rho}
    -\frac{1}{2\rho^2}\frac{\partial\Phi(\rho,z)}{\partial\rho}\right)
    \left(\rho\frac{db(\rho)}{d\rho}-b(\rho)\right)\\
        +\frac{1}{[K(\rho,z)]^3}\frac{\partial\Phi(\rho,z)}{\partial z}
       \frac{\partial K(\rho,z)}{\partial z}+\text{two positive terms}=\\
     \left(-\frac{1}{2\rho^2}\frac{1}{1+\frac{F(z)}{\rho^a}}\frac{aF(z)}
      {\rho^{a+1}}-\frac{1}{2\rho^2}A'(\rho)G(z)\right)
          \left(\rho\frac{db(\rho)}{d\rho}-b(\rho)\right)\\
      +\frac{1}{\left(1+\frac{F(z)}{\rho^a}\right)^3}A(\rho)G'(z)
         \frac{F'(z)}{\rho^a}+\text{two positive terms}. 
\end{multline}
Since $\rho\, db(\rho)/d\rho-b(\rho)$ is negative, the first term on the 
right-hand side of Eq.~(\ref{E:fourth}) is indeed positive.  The second term is 
negative, close to zero near $z=z_1$, but increasing in absolute value as 
$z\rightarrow 0$.  Since $A(\rho)G'(z)$ already occurred in 
Eq.~(\ref{E:chooseAG}), we may have only limited control over this quantity.  
Fortunately, we can adjust the competing term involving $A'(\rho)G(z)$: 
$A(\rho)$ can always be chosen with larger $A'(\rho)$, especially near the
throat, but, more importantly, $\left|G(z)\right|$ can be made as large as we 
please by ``lowering" $G(z)$.  [That $A'(\rho)G(z)$ is relatively large 
was already mentioned after Eq.~(\ref{E:secondcontinued}).]  

The remaining terms have the form
\begin{multline}\label{E:fourthcontinued}
    \left(1-\frac{b(\rho)}{\rho}\right)\left\{A''(\rho)G(z)
           +[A'(\rho)G(z)]^2\phantom{\frac{F(z)/ \rho^a}{(1+F(z)/ \rho^a}}
                 \right.\\ 
               \left. +\frac{1}{\rho}A'(\rho)G(z)\left(1-aL(z)\right)     
                -\frac{a^2}{\rho^2}L(z)\left(1+L(z)\right)\right \},
\end{multline}
where
\[
       L(z)=\frac{\frac{F(z)}{\rho^a}}{1+\frac{F(z)}{\rho^a}}.
\]
At the throat the entire expression is equal to zero.  Moving away from the 
throat, since $A''(\rho)G(z)$ is negative, care must be taken to keep
$A''(\rho)$ small enough so that the entire sum remains positive.

In the last energy condition (\ref{E:lastWEC}),
\[
  \frac{1}{2}(G_{\hat{t}\hat{t}}+G_{\hat{\rho}\hat{\rho}})+\frac{1}{2}
    (G_{\hat{t}\hat{t}}+G_{\hat{z}\hat{z}})
         +G_{\hat{\rho}\hat{z}}\ge0,
\]
the first two terms have already been taken care of.  The last term is zero at 
the throat.  Away from the throat the heavily dominating $A'(\rho)G(z)$
contributed by the first two terms carries the day. 

\section{The upper layers}
So far we have considered only the region from $z=0$ to $z=z_2$ and found
that the WEC need not be violated on this interval.  The next task is to find
a convenient value for $z_2$ at which to cut off and replace the wormhole 
material by a transitional layer without
introducing any new energy condition violation.  The easiest way is to start 
with Eq.~(\ref{E:fourth}) and to assume for the time being that  
$b\equiv 0$, thereby cutting off the wormhole material.  (The transitional
layer will be introduced below.)  The first term on the
right-hand side is zero.  Now choose $z_2$ so that $G'(z_2)$ is small enough 
to keep the right-hand side positive.  Again for the sake of simplicity, 
replace $G$ by a straight line with slope $G'(z_2)$ for $z>z_2$ (or by a 
curve with a very small curvature), thereby retaining continuity.  The line 
will eventually reach zero at some $z=z_3$.  $F(z)$ should also have 
a small curvature for $z>z_2$.  Observe that as $F(z)\to 0$, $K(\rho,z)$ 
approaches unity, so that the metric itself approaches that of a flat 
Minkowski spacetime.  

Since $b\equiv 0$, the ``negative energy expression"
(\ref{negativeenergyexpression}), $\rho \,db(\rho)/d\rho
-b(\rho)$, is also zero, so that the energy conditions are
satisfied \emph{a fortiori}.  [The remaining terms, particularly terms 
involving $1-b(\rho)/\rho$, do not present
any special problems apart from some minor fine tuning: Expressions~
(\ref{E:secondcontinued}) and (\ref{E:fourthcontinued}) suggest that 
$F(z)$ should reach zero before $G(z)$.  Fortunately, $|G(z)|$ is assumed 
to be large to begin with and to decrease relatively slowly.]

The final step is to introduce a transitional layer between $z=z_2$ and 
$z=z_3$: instead of letting $b\equiv 0$, let $b=\epsilon >0$, where 
$\epsilon$ is so small that the resulting ``negative energy expression"
is sufficiently close to zero to leave the above comments unaffected.
According to line element~(\ref{E:line2}), $\rho=\epsilon$ now becomes 
the Schwarzschild radius: since $A(\rho)$ has a vertical asymptote at 
$\rho=\epsilon$, $e^{2\Phi}\rightarrow 0$ as $\rho\rightarrow\epsilon+$ 
for all $z$, thereby creating an event horizon.  So the transitional 
layer is a Schwarzschild spacetime.

\section{The throat}
To complete the discussion, we need to take a closer look at the throat.  So
far we considered only the cylindrical surface $\rho=\rho_0$ from $z=0$ to 
$z=z_2$, the exact analogue of $r=r_0$ for the spherically symmetric case.   
Since the upper layers create a flat top at $z=z_2$, there is no violation of 
the WEC (Visser~\cite{mV96}, chapter 15).  But unlike Visser's cubical 
wormholes, the edges cannot be rounded off, as this would violate the 
condition $\partial b(\rho,z)/\partial z =0$.  It is precisely this rounding 
off that causes the violation of the WEC for cubical wormholes.  

While keeping the sharp edge does solve one problem, it results in another:
by changing the shape function at $z=z_2$ (and again at $z=z_3$), 
$b$ depends on $z$ after all.  In
fact, there is a jump discontinuity at $z=z_2$, so that the partial derivative
with respect to $z$ does not exist.  As a result, the entire solution breaks 
down at this value.  But as noted in the introduction, we are going to make
a small change in one of the assumptions, replacing the regular derivative by
a one-sided derivative: on any closed interval $[z_2,z'_2]$ the right-hand
derivative is
\[
   \frac{\partial b(\rho,z_2+)}{\partial z}=\lim_{z\rightarrow z_2+}
   \frac{b(\rho,z)-b(\rho,z_2)}{z-z_2}=0.
\]
On the open interval $(0,z_2)$ the function is everywhere differentiable with
respect to $z$ and its partial derivative is also equal to zero.  With this
modification we can retain the requirement
\[
     \frac{\partial b(\rho,z)}{\partial z}=0
\]
and hence the earlier analysis.  Observe that the throat is a piecewise 
smooth surface.

While the wormholes described here are not likely to occur naturally, 
a traversable
wormhole that does not violate the WEC could in principle be constructed 
by an advanced cicilization.

\section{Traversability}
Our final topic is a brief look at traversability conditions.  Since our 
variable $\rho$ is analogous to $r$ in the spherical case, we assume that the
traveler approaches the throat along a path perpendicular to the $z$ axis.  This
is best analyzed with the aid of an orthonormal basis in the traveler's frame:
\[
   e_{\hat{0}'}=\mu=\gamma e_{\hat{t}}\mp\gamma\left(\frac{v}{c}\right)
       e_{\hat{\rho}},\,\, e_{\hat{1}'}=\mp\gamma e_{\hat{\rho}}
           +\gamma\left(\frac{v}{c}\right)e_{\hat{t}},\,\,
    e_{\hat{2}'}=e_{\hat{\theta}},\,\, e_{\hat{3}'}=e_{\hat{z}}.
\]
Here $\mu$ is the traveler's four-velocity, while $e_{\hat{1}'}$ points in the 
direction of travel.  (See also Ref.~\cite{MT88}.)

Since $\rho$ is analogous to $r$, the constraint
$\left|R_{\hat{1}'\hat{0}'\hat{1}'\hat{0}'}\right|$ is similar to the spherical
case discussed in MT~\cite{MT88}.

The other constraints show a different behavior.   Thus
\begin{multline}
   \left|R_{\hat{2}'\hat{0}'\hat{2}'\hat{0}'}\right|=
   \left|\gamma^2 R_{\hat{\theta}\hat{t}\hat{\theta}\hat{t}}+
   \gamma^2\left(\frac{v}{c}\right)^2
         R_{\hat{\theta}\hat{\rho}\hat{\theta}\hat{\rho}}\right|=\\
    \left|\gamma^2\frac{\partial\Phi(\rho,z)}{\partial\rho}
    \left(1-\frac{b(\rho)}{\rho}\right)\left(\frac{1}{K(\rho,z)}
        \frac{\partial K(\rho,z)}{\partial\rho}+\frac{1}{\rho}\right)\right.\\
       \left.+\frac{1}{[K(\rho,z)]^3}\frac{\partial\Phi(\rho,z)}
           {\partial z}\frac{\partial K(\rho,z)}{\partial z}\right|\\
     +\gamma^2\left(\frac{v}{c}\right)^2
     \left|\frac{1}{\rho K(\rho,z)}\left(\frac{2\partial K(\rho,z)}
      {\partial\rho}+\rho\frac{\partial^2 K(\rho,z)}
           {\partial\rho^2}\right)
           \left(1-\frac{b(\rho)}{\rho}\right)\right.\\
      \left.-\frac{1}{2\rho^3 K(\rho,z)}\left(\rho\frac{\partial K(\rho,z)}
        {\partial\rho}+K(\rho,z)\right)\left(\rho\frac{db(\rho)}{d\rho}
           -b(\rho)\right)\right|. 
\end{multline}
While the second part is the usual constraint on the velocity of the traveler,
the first part shows that the best place to cross the throat is at $z=z_1$.

\section{Further Discussion}

We have seen that the price to pay for avoiding the violation of the WEC
for a particular type of wormhole is the introduction of a one-sided
derivative.  While it is tempting to argue that mathematically speaking,
the adjustment is rather minor, it does lead to another violation.
Fortunately, this violation is physically defendable: we are merely replacing 
one type of wormhole material by another.  This gives our violation the 
appearance of being less serious than an energy violation. 

Similar questions can be asked about the possible violation of the null
energy condition (NEC) and the averaged null energy condition (ANEC).  
The NEC, $T_{\hat{\alpha}\hat{\beta}}k^{\hat{\alpha}}k^{\hat{\beta}}\ge0$ 
for null vectors, excuses us from 
considering the timelike vector $(1,0,0,0)$, leaving only $\rho+p_j$,
where $p_j$ are the principal pressures.  For the functions considered 
earlier, this condition is met.  The ANEC
is much more problematical.  What is peculiar to our wormhole, however,
is that in the critical non-Schwarzschild region, that is, up to the 
cut-off at $z=z_2$, the quantities $\rho+p_j$ are bounded away from zero.
As a consequence, the ANEC, $\int T_{\hat{\alpha}\hat{\beta}}
k^{\hat{\alpha}}k^{\hat{\beta}}d\lambda\ge0$ (Ref~\cite{mV96}), has a
good chance of being met (although difficult to quantify).  While the
price to be paid is still the same, at least the payoff has enjoyed a boost. 

\section{Conclusion}
This paper discusses traversable wormholes different from those of the 
Morris-Thorne type.  The wormhole geometry is cylindrically symmetric and
the throat a cylindrical surface, a surface that is only piecewise smooth.
As a result, the shape function is not differentiable at some $z=z_2$
due to a jump discontinuity.  
It is proposed that the regular derivative be replaced by a one-sided 
derivative at this value. It is shown that for proper choices of $b$, $\Phi$, 
and $K$ in line element (\ref{E:line2}) the weak energy condition is satisfied 
for all timelike and null vectors.

\end{document}